# Approach to Nuclear Fusion
# Utilizing Dynamics of High-Density Electrons and Neutrals

Alfred Y. Wong* and Chun-Ching Shih

Alpha Ring US Inc., 5 Harris Court, Building B, Monterey, California 93940

*Correspondence author: awong@physics.ucla.edu

8/14/2019

## Abstract

An approach to achieve nuclear fusion utilizing the formation of high densities of electrons and neutrals is described. The profusion of low energy electrons provides high dynamic electric fields that help reduce the Coulomb barrier in nuclear fusion; high-density neutrals provide the stability and reaction rates to achieve break-even fusion where charged particles are the main products. Interactions of energetic charged particles with high-density background produce positive feedbacks with enhanced cross sections. Experiments in a rotating geometry illustrate the advantages of this approach, which discriminates against neutronic fusion.

PACS indices: 52.30.-q; 25.30.-c; 25.70.Jj

## I. Introduction

Conventional nuclear fusion efforts are generally concentrated at overcoming the Coulomb barrier through the confinement of energetic ions of tens to hundreds of keV. This paper describes a different approach through lowering the Coulomb barrier by a profusion of electrons to allow fusion to occur at low energies of tens of eV to several keV and high densities of reactants. New methods of compressing neutrals to high densities in presence of shielding electric fields encourage significant fusion processes to take place through quantum tunneling in a stable environment.

Positive feedbacks, which greatly enhance the fusion process, are made possible in the presence of such high-density operations.

Since the major hurdle in fusion is the Coulomb repulsion between positively charged nuclei, the first feature of our approach consists of using collective temporal and spatial behaviors of free and low-energy electrons to generate negative electric potentials. The Poisson's equation governing potentials depends only on the difference between electron and ion densities and not on their energies. A more favorable potential profile or electric field can be generated by groups of free electrons for quantum tunneling between initially low energy reactants. For example a highly emissive material such as Lanthanum Hexaboride ($LaB_6$) is heated by rotating neutrals to provide a reservoir of free electrons; at the same time the radial movement of electrons is blocked by the same rotating high density ($>10^{26}/m^3$) neutrals at the outer electrode such that the emitted free electrons acquire a high density ($> 10^{22}/m^3$). Oscillatory or collapsing behaviors of these free electrons, such as those existing in resonant plasma waves [1] can further enhance local electron densities. Since fusion events occur in temporal and spatial scales of femtoseconds and femtometers, even high frequency dynamic electric fields can counteract the repulsive Coulomb electric field between reactant nuclei, thereby causing fusion.

The second feature is to increase the number of reactants through the emphasis on fusion between neutrals, which do not cause instabilities. A method of controlling the motion of neutrals through a minority group of ions was developed from the strong coupling between high-density neutrals and ions through collisions [2-6]. A highly compressed state including neutrals, electrons and ions through centrifugal or ponderomotive forces results in significant fusion events since the fusion rate depends on the product of densities of reactants.

The third feature is to encourage positive feedbacks among fusion products and high-density reactants; the mutual enhancement between the population of free electrons and charged MeV fusion particles results in larger cross sections above a certain threshold that is related to the generation of electrons and ions in high density neutrals and solid emitters. This feedback process is found to favor aneutronic fusion with charged particles as products and discriminate against neutronic fusion whose neutrons do not interact with the immediate environments.

In this paper we aim to demonstrate that as the Coulomb barrier is reduced by localized electric fields, cross sections for p-B fusion improve from vanishingly small to finite ($10^{-37}$ m$^2$) at low energies (tens of eV). High neutral densities of reactants, which do not give rise to plasma instabilities, are present to yield significant fusion events, since the rate of fusion is proportional to the product of reactant densities. Experimental results are presented to demonstrate that above a threshold, given by the energy required to achieve electron and neutral dynamics, exponential growths of fusion events occur. Particles of MeV energy are produced as evidenced by tracks in CR 39 and chamber wall. A net gain in the output power is observed through recordings of temperature changes in the internal structures and coolants. Optical signatures of new elements are also recorded.

Although this approach was first discovered in experimental settings, we have chosen, for pedagogical reasons, to start with a theoretical account. A general discussion of how negative electric fields affect the fusion cross section is given in Section 2. The screening energies as the result of dynamic screening by electron distributions are discussed in Section 3. The energetic particles produced from fusion reactions generally will raise the temperature and ionization of background neutrals, which in turn increase the emission of electrons, in favor of enhancing the fusion cross section. This process of positive feedbacks will be described in Section 4. An experiment using a rotating geometry to illustrate the foregoing principles is discussed in Section 5. The experimental support of our concept is shown in Section 6. The generality of our approach to fusion processes is presented in the concluding Section 7.

## II. Effects of electric potential on cross section of fusion interaction

One of the key factors in calculating fusion reaction rates is the cross section, which is usually very small at low energy due to the extremely low penetration factor through the Coulomb barrier around the nucleus. The tunneling through the Coulomb barrier is purely a quantum mechanical phenomenon and can be described by the Coulomb scattering process based on the Schrodinger equation,

$$-\frac{\hbar^2}{2m}\nabla^2\psi + [V(r) - E]\psi = 0 \tag{1}$$

where $V(r) = Z_1Z_2e^2/r$ is the Coulomb potential between particles of charges $Z_1e$ and $Z_2e$. Eq. (1) can be solved for the Coulomb wave function around the nucleus [7]. Since the nuclear radius (~$10^{-15}$m) is much smaller than the Coulomb radius (~$10^{-10}$m), the penetration probability $P$ is related closely to the wave function near $r = 0$ and obtained as,

$$P = \frac{2\pi\eta}{\exp(2\pi\eta)-1} \quad ; \quad \eta = \frac{\alpha Z_1 Z_2 c}{v} \tag{2}$$

where $\alpha = e^2/\hbar c$ is the fine structure constant, v is the particle velocity, and $c$ is the light velocity in vacuum. The constant $\eta$ can be expressed in terms of the particle energy $E$ [8],

$$2\eta = \sqrt{\frac{E_G}{E}} \quad ; \quad E_G = 2\alpha^2 Z_1^2 Z_2^2 mc^2 \tag{3}$$

where $E_G$ is the Gamow energy equivalent to the Coulomb potential at the Bohr nuclear radius. $E_G$ is proportional to the nuclear charges squared, making it much more difficult to get fusion going between high-Z nuclei. Since the reaction rate is proportional to the penetration factor, the fusion cross section is customarily written as [8]

$$\sigma(E) \sim \frac{P(E)}{v} = \frac{S(E)}{E} \frac{1}{\exp\left(\pi\sqrt{\frac{E_G}{E}}\right)-1} \tag{4}$$

where $S(E)$ is the astrophysical factor, which represents a modification of actual cross sections from the ideal formula in (2). Usually $S(E)$ has a weak dependence on $E$ and can be expanded in power of $E^n$ where the coefficients of expansion are determined from experiments.

When the penetration probability is small, the fusion cross section can also be calculated from the WKB method. Under this approximation, the barrier penetration factor $P$ can be found as

$$P = \exp\left[-\frac{2\sqrt{2m}}{\hbar} \int_{r_1}^{r_2} \sqrt{V(r)-E}\, dr\right] \tag{5}$$

where the integral covers the radial range where the argument of the square root is positive, i.e. $E < V(r)$. For unscreened Coulomb potential, $V(r) = Z_1Z_2e^2/r$, the integral can be carried out exactly and we have

$$P = \exp\left\{-\frac{2\pi Z_1 Z_2 e^2}{\hbar v}\left[1-\frac{2}{\pi}\left(\sin^{-1}\sqrt{\xi}+\sqrt{\xi-\xi^2}\right)\right]\right\} \quad \text{where} \quad \xi = \frac{E}{U_{pk}} \tag{6}$$

and $U_{pk}$ is the peak potential barrier. In the case of small $E$, $\xi$ is negligible and

$$P \approx \exp\left(-\frac{2\pi Z_1 Z_2 e^2}{\hbar v}\right) = \exp\left(-\pi\sqrt{\frac{E_G}{E}}\right) \tag{7}$$

which has exactly the same exponential dependence on energy $E$ as given in (2) when $E$ is much smaller than $E_G$ (a generally valid approximation for most cases).

If an external negative potential is created besides the positive Coulomb potential, the *height and width* of overall Coulomb barrier to be penetrated will become smaller. Usually, this external potential comes from the electrons around the nuclei and typically has a characteristic distance longer than the separation between nuclei; as shown below this property is of great importance to quantum tunneling and therefore to fusion. Generally it can be considered that the Coulomb potential is reduced by a constant value, $E_s$, called the screening energy. Including this effect, the Schrodinger equation becomes

$$-\frac{\hbar^2}{2m}\nabla^2\psi + [V(r) - E_s - E]\psi = 0 \tag{8}$$

Equation (8) is identical to (1) if $E$ in (1) is replaced with $(E + E_s)$. As a result, we can use the same formula as in (2) and obtain the penetration probability with screening effects as

$$P = \pi\sqrt{\frac{E_G}{E+E_s}}\left[\exp\left(\pi\sqrt{\frac{E_G}{E+E_s}}\right) - 1\right]^{-1} \tag{9}$$

And the screened cross section, like (4), can be written as

$$\sigma(E, E_s) = \frac{S(E)}{E+E_s}\left[\exp\left(\pi\sqrt{\frac{E_G}{E+E_s}}\right) - 1\right]^{-1} \tag{10}$$

where we assume the astrophysical factor remains the same under the screening effect.

The screening energy can thus affect the fusion cross section, especially when it is close to or higher than the particle energy. Figure 1 shows the cross sections for p-$^{11}$B reaction screened with electron densities up to 6x10$^{25}$/m$^3$ ($E_s \sim$ 27.6 keV assuming $\Lambda = 1\mu$m) using (10) along with $E_G =$ 2.29 MeV and the following empirical astrophysical factor [9]

$$S(E) = 195 + 247\,E + 231E^2 \tag{11}$$

where $S(E)$ is in MeV-barn and $E$ in MeV.

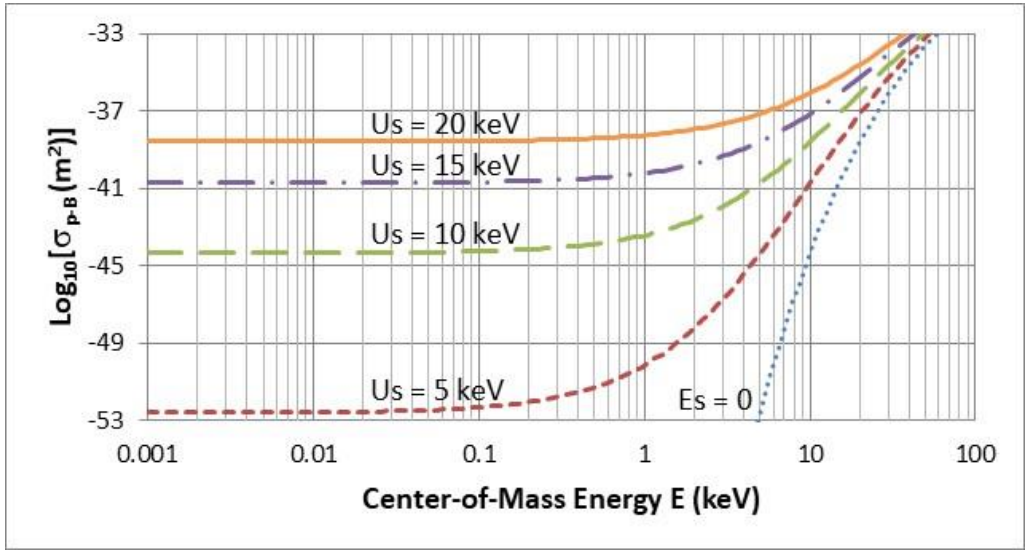

Figure 1: p-$^{11}$B cross section as function of particle energy for the screening electron densities up to $E_s = 20$keV. The cross section near $E = 10$eV grows over 14 orders of magnitude (from $10^{-53}$ to $10^{-39}$m$^2$) over the range of 5 to 20keV.

Without screening ($E_s = 0$), the cross sections drop quickly when the particle energy goes below a few tens of keV. With the screening effect, the cross sections turn nearly flat below a few keVs of $E$ because the screening energy dominates over the particle energy in this region. Slightly elevated cross sections as $E \rightarrow 0$ are due to the appearance of $E^{1/2}$ in the denominator of $\sigma$. It is clear the cross section rises very quickly with the screening energy (from $10^{-53}$ to $\sim 10^{-35}$m$^2$ as $n_1$ rises from $1 \times 10^{25}$ to $6 \times 10^{25}$/m$^3$). The strong ion-neutral coupling is responsible in accelerating neutrals to energies around 20-40 eV (Appendix C). In the following section, we'll show how to create a fusion environment with high screening energies and compare the effect to other known processes, such as muon-catalyzed fusion [10, 11].

**III. Calculation of screening energies**

In the previous section, we have shown that screening energy is a key factor in quantifying the screened cross section. It is possible to create a negative potential with the manipulation of electron density distributions around the positive Coulomb potential and determine the effect by calculating the corresponding screening energy. It is important to point out what really causes the screening effect is the electric field generated by collective electron motions. Electrons do not have to be located physically between two fusing nuclei. In the following we consider fusion in ion-neutral and neutral-neutral interactions screened by bound electrons or dynamically free electrons. Results are also compared with muon-catalyzed fusion reactions.

*Hydrogen atom approached by a proton*

The simplest case is a hydrogen atom target, which consists of an orbiting electron around a proton. We'll calculate its screening energy and extend to other cases of interest.

Assume the hydrogen atom is at ground state. The electron wave function is well known to be [7]

$$\psi(r) = \frac{1}{\sqrt{\pi}a_0^{2/3}} \exp\left(-\frac{r}{a_0}\right) \quad ; \quad a_0 = \frac{\hbar^2}{me^2} \tag{12}$$

where $a_0$ is the Bohr radius for hydrogen atom. The electric potential inside the atom can be found from the Poisson's equation $\nabla^2 V(r) = 4\pi e |\psi(r)|^2$ and the result is

$$V(r) = \frac{e^2}{r}\left(1 + \frac{r}{a_0}\right) \exp\left(-\frac{2r}{a_0}\right) \tag{13}$$

For small $r$ or large $a_0$, the potential approaches

$$V(r \to 0) \approx \frac{e^2}{r} - \frac{e^2}{a_0} \equiv \frac{e^2}{r} - E_s \tag{14}$$

Therefore, we found the screening energy $E_s$ for a hydrogen atom is equal to the Hartree energy, defined as $E_H = e^2/a_0 \sim 27.2\text{eV}$. Eq. (14) was derived based on the assumption that the incoming proton will not affect the electron spatial distribution. However, we know that, as the proton moves closer to the target hydrogen nucleus, the total charge is doubled, and the electron orbit will be pulled in and could double the screening energy. So, the overall effective screening energy for hydrogen atom impacted with a proton (ion on neutral) should be somewhere between $E_H$ and $2E_H$.

Since $E_s$ is proportional to $Z^2/m$, we calculate the screening energy for boron atom to be $E_s$(boron) = 371 eV, consistent with the measured value of 430±80 eV [12,13].

*Muon-hydrogen screening effect*

Muon-hydrogen atom is formed when the electron is replaced with a muon. The formulas for the ordinary hydrogen atom can be applied to the muon-hydrogen, except the electron mass needs to be replaced with the muon mass, which is about 207 times heavier. Therefore, we expect the screening energy for the muon-hydrogen with an incoming proton is

$$E_{s,\mu} = E_{s,e} \frac{m_\mu}{m_e} \approx 5.63 \text{ keV} \tag{15}$$

Eq. (15) shows that the screening energy is proportional to the mass of the orbiting lepton.

*Screening between two approaching hydrogen atoms*

In most cases, we are interested in two neutral atoms approaching each other, with screening electrons around each nucleus. The electric potential between two nuclei as function of the distance can still be derived from the Poisson's equation for the two approaching hydrogen atoms and the resulting potential is (Appendix A)

$$V(r) = \frac{e^2}{r}\left(1 + \frac{5r}{8a_0} - \frac{3r^2}{4a_0^2} - \frac{r^3}{6a_0^3}\right)\exp\left(-\frac{2r}{a_0}\right) \tag{16}$$

and $V(r \to 0) \to \frac{e^2}{r} - \frac{11}{8}\frac{e^2}{a_0}$  thus  $E_s = \frac{11}{8}E_H$

Therefore, the neutral-neutral screening energy is greater than the ion-neutral case by about 40 percent. As pointed out earlier, when the two neutrals move to nearly overlap with each other, both inner orbiting electrons will be subject to the charge of two nuclei and the screening energy can move closer to $2E_H$.

*Dynamic electron oscillation screening*

It is obvious that a uniform electron density cannot generate needed screening energy because there is no variation in the electric potential. For high-density electron plasmas in a matching positive charge background, such as in metals, the best way to generate high electric fields is by resonant plasma oscillations or plasma waves. The fusion process has small spatial and temporal scales, which are shorter than those of plasma oscillations in the visible range, such as plasma resonances on the surface of metals. The groupings of electrons in the plasma oscillation could produce high oscillating electric fields to offset the Coulomb repulsion between fusing nuclei.

Assume the plasma wave has the sinusoidal form of

$$n(x) = n_0 + n_1 \cos\left(\frac{2\pi x}{\Lambda}\right) \tag{17}$$

where $n_1$ is the plasma wave amplitude and $\Lambda$ is the wavelength. From the Poisson's equation, the corresponding electric potential energy can be found to be

$$V(x) = \frac{n_1 e^2 \Lambda^2}{\pi} \cos\left(\frac{2\pi x}{\Lambda}\right) \tag{18}$$

The amplitude of the potential variation is thus the screening energy, which can be written as

$$E_s = \frac{n_1 e^2 \Lambda^2}{\pi} \approx 4.6 \times 10^{-10} n_1 \Lambda^2 \text{ (eV)} \tag{19}$$

For a typical plasma wave induced by visible or infrared laser, the wavelength is $\Lambda \sim 10^{-6}$ m. In this case, the screening potential is $E_s = 4.6$ keV for a plasma wave amplitude of $n_1 = 10^{25}$ m$^{-3}$. This energy is comparable to the screening energy for the muon-catalyzed hydrogen fusion (5.6 keV). Therefore, we expect an electron plasma wave of amplitude ~1.2x10$^{25}$ m$^{-3}$ around a hydrogen atom would result in a cross section similar to the muon-catalyzed fusion. Apparently, such plasma waves should be easier to produce than the short-lived muons. The nonlinear collapse [1, 14] of plasma waves can lead to spatial scales of a tenth of plasma wavelength. These highly localized oscillating electric fields lead to ponderomotive forces that are proportional to the gradient of field intensities. These forces compress positive ions together and help to lower the Coulomb barrier between positively charged nuclei.

*Long-wavelength electron screening*

Since the shielding electric field $\underline{E}$ is proportional to the surface charge density ($\underline{E} \sim nl$), the screening energy thus behaves as $E_s \sim \underline{E}l \sim nl^2$ where $l$ is the characteristic length of such electron density slab, as shown in Eq. (16). The required electron density fluctuation can thus be greatly relaxed for the same screening effect. Such a situation can exist near the surface of electron emitters as discussed in a later section on plasma fusion echoes.

**IV. Enhanced fusion with positive feedback**

The key feature of this new fusion concept depends on the screening effect of electrons around the neutrals. It is expected that the fusion process will release more electrons through heating or collisions with fusion products. These processes could cause larger electron density fluctuations, such as Langmuir collapses [1, 15]. This type of positive feedback generates stronger screening effects and could create sustainable fusion process for energy production (Figure 2).

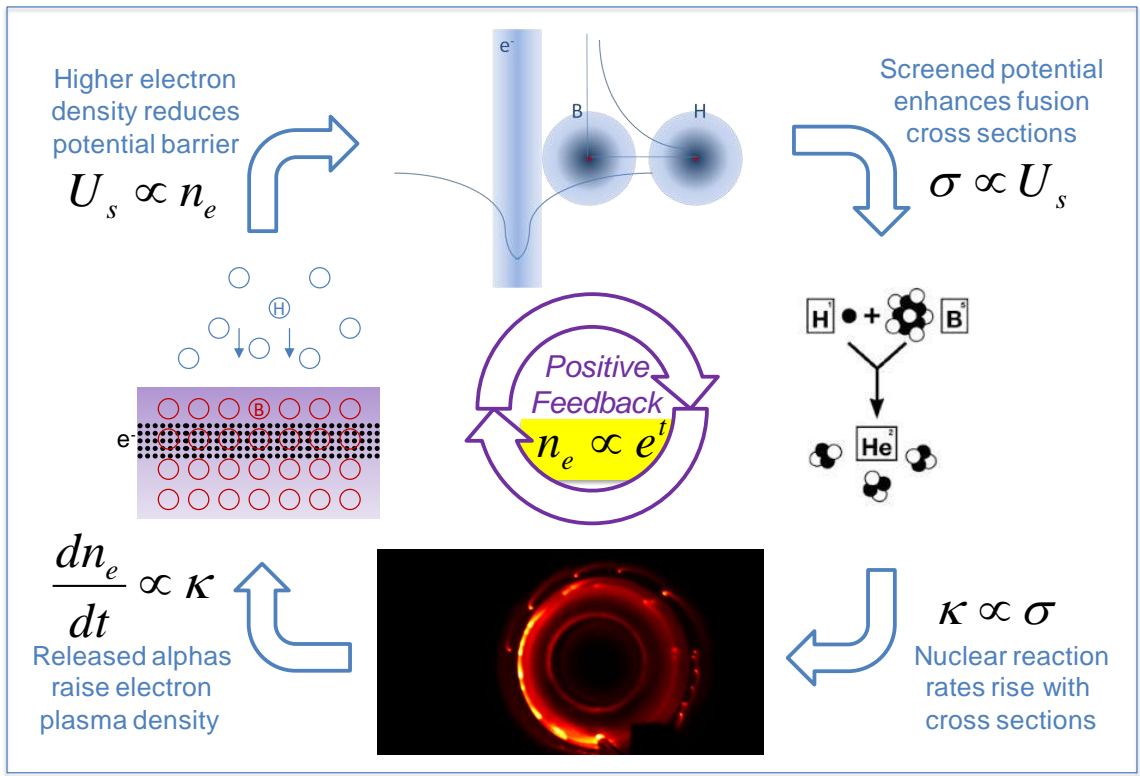

Figure 2: Positive Feedback in which energetic charge particles from fusion raise the temperature of electron emitters to produce more free electrons, which in turn enhance cross sections and thus the fusion process. The same charge particles also ionize background neutrals to increase the population of free electrons.

For example, three high-energy (MeV) alpha particles are produced from each p-$^{11}$B reaction. Each alpha particle could induce about $10^5$ electrons in its path before its energy is completely dissipated [16-18]. These electrons thus provide effective screening mechanism to further the fusion process. The change of the electron density fluctuation can be written as

$$\frac{dn_1}{dt} = \beta\gamma\frac{dn_\alpha}{dt} = 3\beta\gamma\sigma v n_H n_B = 3\beta\gamma\kappa \tag{20}$$

where "3" comes from "3" alpha particles per reaction, $\beta$ (~ $10^5$) is the number of electrons per alpha, $\gamma$ (~ 0.01) is the fractional plasma oscillation amplitude, $\sigma$ is the cross section, v is the relative velocity between hydrogen and boron, $\kappa$ is the reaction rate, and $n_\alpha, n_H, n_B$ are the densities of alpha, hydrogen, boron.

In general, if the electron density does not change much from the initial value during the fusion process, the $E_s$-dependent cross section can be expanded as

$$\sigma(E_s) \approx \sigma(E_{s0}) + \sigma'(E_{s0})(E_s - E_{s0}) + \cdots \tag{21}$$

where $E_{s0}$ is the initial screening energy and $\sigma'$ is the derivative of $\sigma$ with respect to $E_s$.

From (19), (20), and (21), we can write an equation for $E_s(t)$ as

$$\frac{dE_s}{dt} = b\left(E_s - E_{s0} + \frac{\sigma}{\sigma'}\right) \; ; \; b = \frac{3\beta\gamma e^2 \Lambda^2 v n_H n_B \sigma'}{\pi} \tag{22}$$

where constants $b$ is proportional to the slope of cross section $\sigma'(E_{s0})$. Equation (22) can be solved easily for time-dependent $E_s$,

$$E_s = E_{s0}[1 + q(e^{bt} - 1)] \; ; \; q = \frac{\sigma(E_{s0})}{\sigma'(E_{s0})E_{s0}} \tag{23}$$

Since the screening energy is proportional to the electron density fluctuation amplitude, we can also write the time-dependent amplitude as

$$n_1(t) = n_1(0)[1 + q(e^{bt} - 1)] \tag{24}$$

Note that $b$ is the temporal rate of exponential growth and $q$ can be considered as a growth factor. It can be seen from (23), $q$ is a dimensionless quantity depending only on the intrinsic property of the particular fusion reaction. Its dependence on the screening energy $E_s$ is shown in Figure 3 for the p-B reaction. At $E_s = 20$ keV, the growth factor $q$ is approximately 0.03.

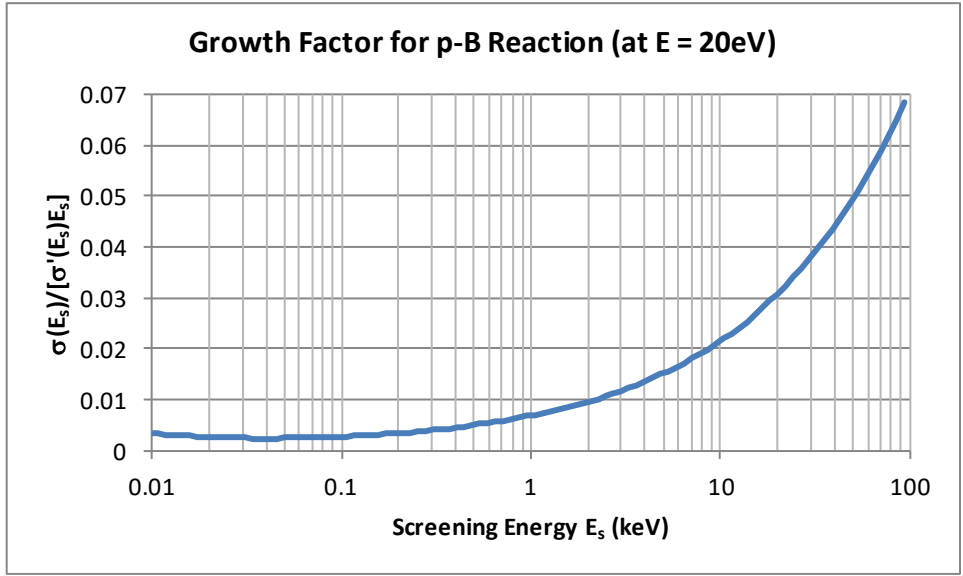

Figure 3: Growth factor $q = \sigma(E_s)/[\sigma'(E_s) E_s]$ as function of $E_s$ for p-B reaction at $E = 20$eV.

Eq. (24) describes initial temporal behaviors of the fusion process. However, there are other mechanisms, such as damping of plasma waves, which might dissipate the exponential growth. As a result, the process exhibits a threshold phenomenon. The output power grows only when the fusion growth overcomes the damping of plasma waves. On the other hand, the fusion-induced plasma fluctuations cannot grow indefinitely. The fractional plasma density fluctuation amplitude certainly may not grow as fast as the average electron density. Therefore, the power growth will be eventually saturated by limited fluctuations. These two phenomena, threshold and saturation, are formulated and discussed further in Appendix B.

The growth rate of density fluctuation amplitude or screening energy, $b$, can be calculated as

$$b = \frac{3\beta\gamma e^2 \Lambda^2 v n_H n_B \sigma'}{\pi} = \frac{3\kappa E_H \beta \gamma a_0 \Lambda^2}{\pi q E_s} \approx 1.33 \times 10^2 \text{ s}^{-1} \tag{25}$$

Eq. (25) shows that the characteristic time of the exponential growth is in the range of a few milliseconds, consistent with what was observed in the lab.

A rough estimate of the expected output power can be obtained from our experimental conditions. The output power per unit volume can be calculated from

$$\frac{P}{V} = Q\kappa = Q\sigma v n_p n_B \tag{26}$$

where $Q$ is the energy released per p-B reaction and $\kappa$ is the reaction rate per unit volume. Using the values of $Q$ = 8.7 MeV, $\sigma$ = $10^{-37}$m², v = 6.3x10⁴m/sec, $n_p$ = $10^{26}$m⁻³ and $n_B$ = $10^{29}$m⁻³, we obtain the reaction rate per unit volume as

$$\kappa = \sigma v n_p n_B \approx 6.3 \times 10^{22} \text{ m}^{-3}\text{s}^{-1} \tag{27}$$

and the output power density as

$$\frac{P}{V} \approx 8.77 \times 10^{10} \frac{\text{W}}{\text{m}^3} = 87.7 \frac{\text{kW}}{\text{cm}^3} \tag{28}$$

This value is consistent with the output of 20 kW measured in an experiment with a reaction volume of approximately $V \sim 0.25$ cm³.

To derive the dependence of output energy on input energy, we need to understand how these two quantities are related to the parameters in (24). The output power can be obtained from (20) and (26),

$$\frac{dE_{out}}{dt} = P_{out} = VQ\kappa = \frac{VQ}{3\beta\gamma}\frac{dn_1}{dt} \tag{29}$$

Therefore, the output energy density can be written as

$$\frac{E_{out}}{V} = \frac{Q}{3\beta\gamma}(n_1 - n_{10}) = \frac{Q}{3\beta\gamma}\frac{\pi}{e^2\Lambda^2}\frac{\sigma}{\sigma'}(e^{bt} - 1) \tag{30}$$

where the relation between $E_s$ and $n_1$ in (19) has been used. In the first-order approximation, the growth rate $b$ is expected to be proportional to the input power density. Therefore, Eq. (30) can be re-written as

$$\frac{E_{out}}{V} = q'\left[\exp\left(b't\frac{P_{in}}{V}\right) - 1\right] \tag{31}$$

$$\text{where} \quad q' = \left(\frac{Q}{3\beta\gamma}\frac{\pi}{a_0\Lambda^2}\frac{E_{s0}}{E_H}\right)q \quad \text{and} \quad b' = \frac{b}{P_{in}/V} \tag{32}$$

are constants related to $q$ and $b$ in (22) and other parameters. From (25) and our experimental conditions, we estimated that $b' \approx 7.8 \times 10^{-8} \text{m}^3/\text{J}$. This dependence is used in a comparison with the lab observation in Section VI (Figure 6).

Sometimes it's more convenient to express the reaction rate, κ, in terms of the center-of-mass energy $E$, instead of the relative velocity v. Using (10) in (26), we have

$$\kappa = \sigma \phi_p n_B = \frac{S \phi_p n_B}{E + E_s} \left[ \exp\left(\pi \sqrt{\frac{E_G}{E+E_s}}\right) - 1 \right]^{-1} \quad (33)$$

Using $E = \mu v^2/2$ where $\mu = m_p m_B/(m_p + m_B)$ is the reduced mass between two interacting atoms (e.g. proton and boron), Eq. (33) can be written as

$$\kappa = \sigma \phi_p n_B = \frac{4 S \phi_p n_B}{E_G} \frac{\eta^2}{e^{2\pi\eta} - 1} \quad \text{where} \quad 2\eta = \sqrt{\frac{E_G}{E+E_s}}, \quad (34)$$

and $\eta$ is the Sommerfeld factor. It is shown in (34) that the reaction rate depends only on the combination of $E$ and $E_s$, not individually. Therefore, higher $E$ would reduce the requirement on the screening energy. Although $E$ is about 20 to 40 eV in the current experiment, we are ready to raise the energy to a few keV through imposing higher $\underline{j} \times \underline{B}$ forces (Appendix C). As a result, the system would require less screening energy to attain similar reaction rates.

## V. Rotating system to illustrate feasibility of fusion

Our laboratory has designed an experiment that encompasses the three features described previously: profusion of oscillating electrons, gas compression with high neutral density and reaction enhancement with positive feedbacks.

First a special method is invented to rotate a large number of neutrals by a small number of ions (2-5 orders smaller), through frequent collisions. No mechanical rotors were used. The fluid equations governing this behavior of strong ion-neutral coupling are shown in Appendix C.

With hydrogen pressures in the torr range ($n \sim 10^{23} \text{ m}^{-3}$), where the mean free path is in microns, this strong coupling causes ions and neutrals to move together. This method has been validated in laboratory experiments as shown in Appendix C. At a rotation rate of $\omega/2\pi = 10^5$ revolutions/s, the neutral density at the outer edge was increased from the initial density of $n_0$ by a factor, $[\alpha\, e^\alpha / (e^\alpha - 1)]$ ~1000, where $\alpha = (m\omega^2 r^2/2kT)$, to $10^{20}$ cm$^{-3}$ [Appendix D]. This large neutral density was confirmed experimentally from a piezoelectric pressure gauge (Prosense SPT25) mounted at the outer electrode. This large neutral density is consistent with the amount of H alpha emission at 656 nm. The high rotation rate of $10^5$ RPS was measured by a fast camera (FASTCAM SA5) capable with a sampling speed of 775,000 frames per second.

We have chosen a rotating system because it is desirable to have fusion at the outer edge of the chamber where electrons, neutrals, ions are all compressed by a large centrifugal acceleration (~$10^{10}$ m/s$^2$, a billion times stronger than the gravitational acceleration on Earth). The outer edge also offers advantages of easy access to measurements and the extraction of energy.

Our experiments are performed in a cylindrical system with two concentric electrodes immersed in an axial magnetic field $B$ (~ 0.1-1.0 T). The experimental chamber is filled with 100% hydrogen gas up to 3.5 torr. Two operating modes have been attempted: a short-pulse mode with a capacitor bank and a long-pulse/steady-state mode with a DC supply.

In the pulse mode, after a short initial pulse (1.4 kV, 0.5 ms) of ionization, a radial current I (~3 kA) is caused to flow between these electrodes (diameters of 9 and 16 cm respectively), by an imposed voltage of 0.5 kV, resulting in an azimuthal rotation driven by $I\,\underline{L} \times \underline{B}$ forces, where $\underline{L}$ is the radial path of the current (3.5 cm).

This current, carried by a radial ion current, flows outward along most of the radial distance between inner (positively biased) and outer electrodes (negatively biased) as ions have greater mobility (by $M_i / m_e$). This ion current is continued onto a radial electron current flowing inward from the outer boundary where the source of electrons comes from an anchored emitting LaB$_6$. This LaB$_6$ piece was heated to near 1900K by the rotation of hydrogen gas as measured by a thermocouple, calibrated with a pyrometer. This electron emission was further enhanced when bombarded by MeV particles during fusion events.

Fusion is observed in a short pulse (12 msec) experiment to occur with the following diagnostics: CR39 plastic detectors register energetic MeV particles from fusion processes; oscillating electric fields for screening effects are demonstrated by the appearance of fusion echoes at discrete azimuthal locations.

In the long-pulse/steady-state mode, the system is run at 200 – 500 V up to 30 A of current after an initial discharge pulse. The system, cooled by circulating water, is specially suited for measurements of heat for consideration of energy balance.

**Fusion echoes due to electrons carried by azimuthal neutral flows**

As shown in Figure 4, a fast camera with two optical filters centered at 589 nm with a width of 7 nm records images of the cross section of our experimental device. The fast camera was able to record the entire duration 12.85 msec with a resolution of 0.01 msec. The optical emissions from helium neutrals and carbon ions, products of p-$^{11}$B fusion, fall into this optical spectral range.

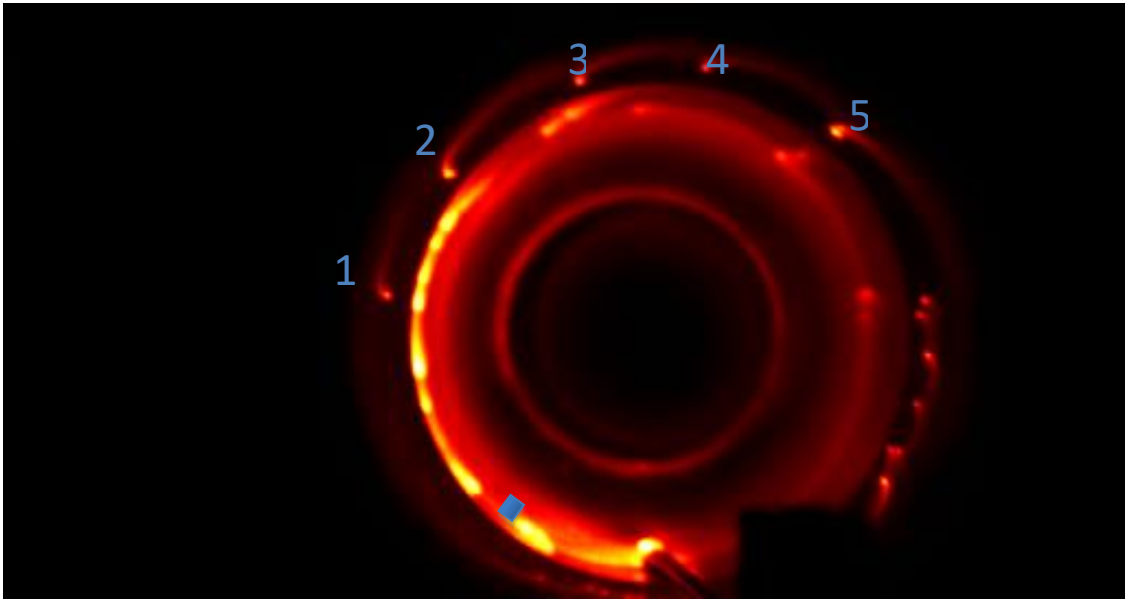

Figure 4: Cross-section showing emission from He I neutrals and CII carbon ions using double filters with center wavelength at 589nm and bandwidth of 7 nm. One LaB$_6$ electron emitter is at the 6 o'clock (or 30-minute) position. The second emitter, BN source, marked by a blue stub is at 37-minute position. The echo positions due to both sources are marked by numerals 1-5.

Charged particles, $^{12}C^+$ ions are predominantly emitted at certain azimuthal angles where fusion locations can be observed more easily against a dark background (Figure 4). Because these energetic particles have low collisional cross sections with background neutrals, their orbits are mostly influenced by the axial magnetic field. Their cyclotron radii are equal to the diameter of the outer electrode because particles with larger orbits cannot escape through the aperture of the outer electrode/shroud.

Echo-like processes observed in Figure 4 are generated from two emitters which are sources of electron emission: one $LaB_6$ source at 30-minute clock-position and one BN emitter at the 37-minute clock-position around the periphery of the outer electrode.

For echoes to occur there must be oscillations of electron densities at each source, which give rise to time-varying electric fields. Past experiments have shown that counter flows of ions and electrons are found to create oscillating double charge layers [19, 20].

As such oscillating electrons are carried in the azimuthal direction by the rotating neutrals, fusion processes occur when propagating electron densities from different sources arrive in phase at specific locations. Fusion process is identified by the emission of energetic ions whose orbits are influenced by the axial magnetic field. These "fusion echoes" demonstrate that oscillating electrons play important roles in the fusion process through creating high densities and fields and that there is a threshold for the reduction of the Coulomb barrier. The locations of echoes depend only on the distance between the two sources and the ratio of the oscillation frequencies of these two sources [21, 22]. A short description of the theory of plasma echoes is given in Appendix E.

**Detection of energetic MeV particles from fusion**

*Detection by CR39 plastic detectors*

Particles of MeV energies generated from fusion are measured with CR-39 plastic detectors [23] made from a hydrocarbon compound (Columbia Resin 39 orpoly allyl diglycol carbonate). Such energetic particles damage the chemical bonds in their penetration through the detector. This commonly used diagnostic for energetic particles does not require any elaborate electronics. When exposed to a 6.25M NaOH solution for 75 minutes, the damaged portion of the detector is etched more quickly than the undamaged portion. The diameters and contrasts (related to depths of penetration) of the resulting tracks are measurable and can be compared with those simulated using the computer modeling software called TRACK_TEST [24]. This detection method was calibrated with an Am241 source whose emitted alpha energies are controlled by the thickness of aluminum sheets in front of the Am241 source and calibrated by a solid-state detector (ORTEC U-015-100-100 Ion-Implanted Silicon Charged Particle Detector). The tracks from experiments and an Am241 source of similar energies are compared as shown in Figures 5a-5d. Our measurements of various diameters and contrasts of tracks yield an estimate of energy in the range of 3-5 MeV for particles produced from fusion. The contrast is defined as the ratio of dark portion to the lighted portion in the image of a track. The depth of the track inside the CR39 is related to this contrast. The diameter and contrast together determine the kind of particles to be detected [23].

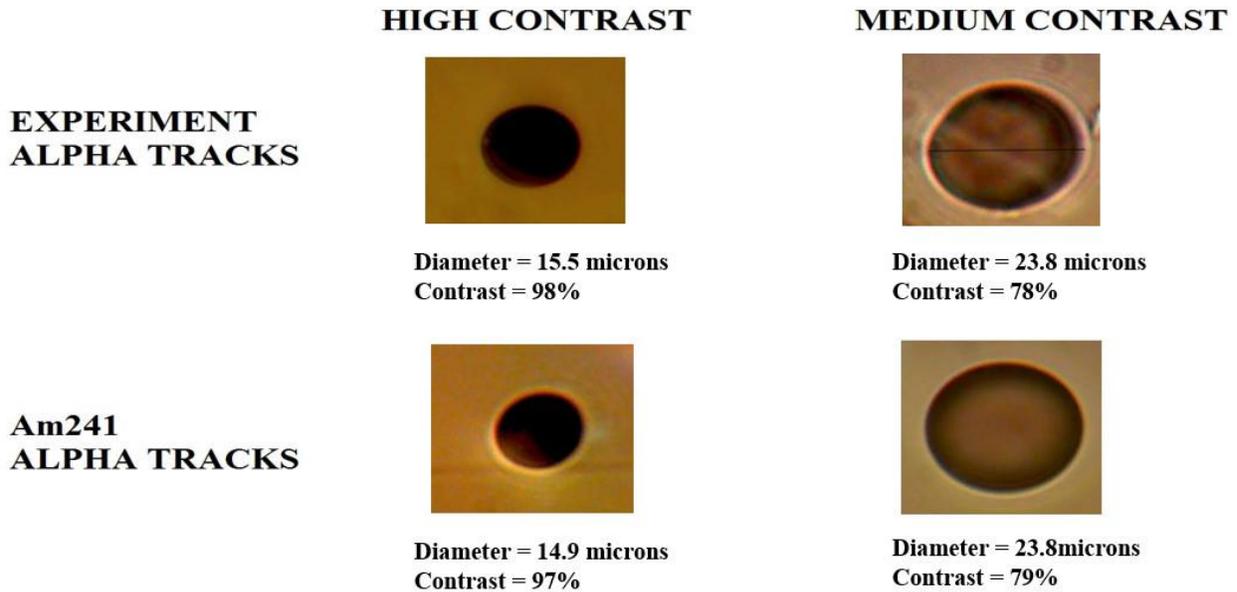

Figure 5a: Comparison between experiment and calibration of alpha tracks on CR39 plates.

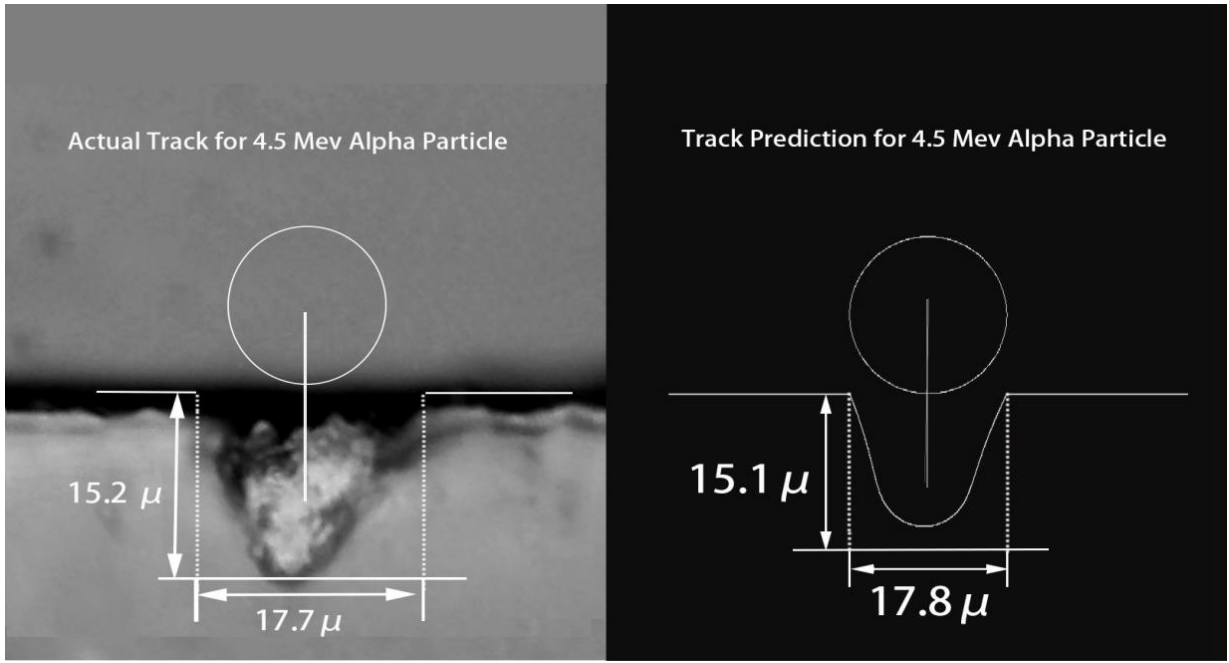

Figure 5b: Comparison between Alpha Track observed under microscope and Alpha Track as predicted by the Track Test Software.

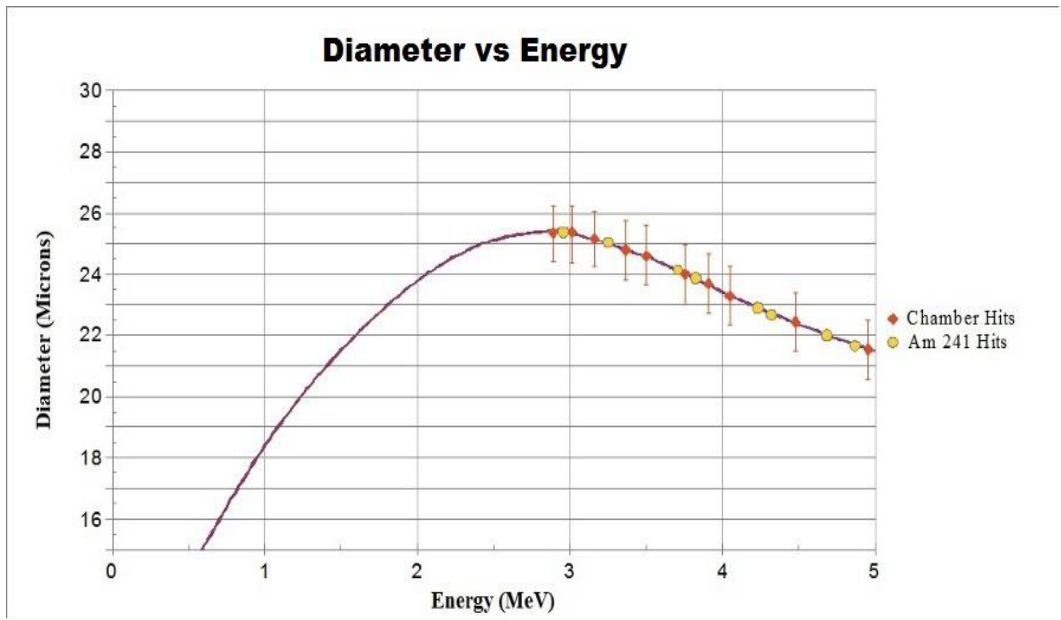

Figure 5c: Diameters of tracks in CR39 detectors vs. energy calibrated by known energy of alphas from Am241 source. Curve is from computer modeling of alphas on CR39. The placement of the data points is determined from their contrasts as shown in Figure 5d.

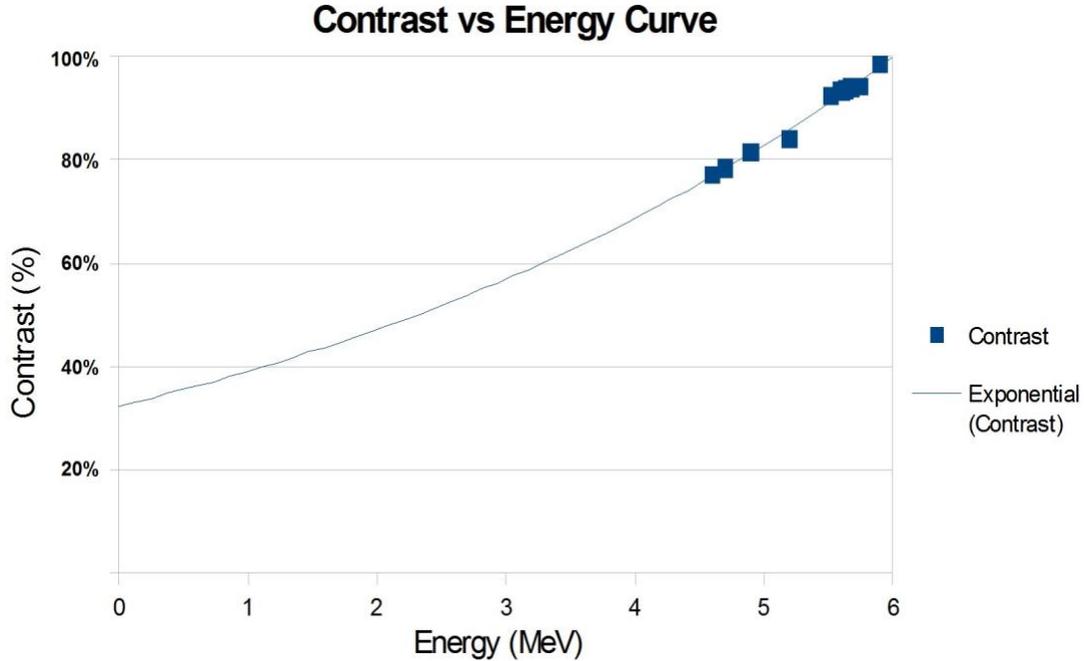

Figure 5d: Contrasts of tracks of experimentally observed energetic particles vs. particle energy.

## VI. Experimental support of new approach.

We have already shown in the previous section that MeV particles are produced even though neutrals and electrons are in the 20eV range. We have also observed the optical emission of new elements such as helium and carbon, which can only appear as the products of p-$^{11}$B reaction. We will comment next on how experiments have provided further support for fusion processes.

**Positive feedback**

The rate of fusion is characterized by positive feedbacks in which the electron population and the fusion product in charge particles enhance each other. As shown in the theoretical treatment, the fusion product in terms of MeV particles should undergo an exponential rise. We have plotted the percentage increase of MeV particles (measured by CR 39) as function of the input power in Figure

6. There is a threshold followed by an exponential rise. The measured growth coefficient $b' \sim 8.2 \times 10^{-8}$ m$^3$/J (eq 32) is close to the theoretical value of $7.8 \times 10^{-8}$ m$^3$/J.

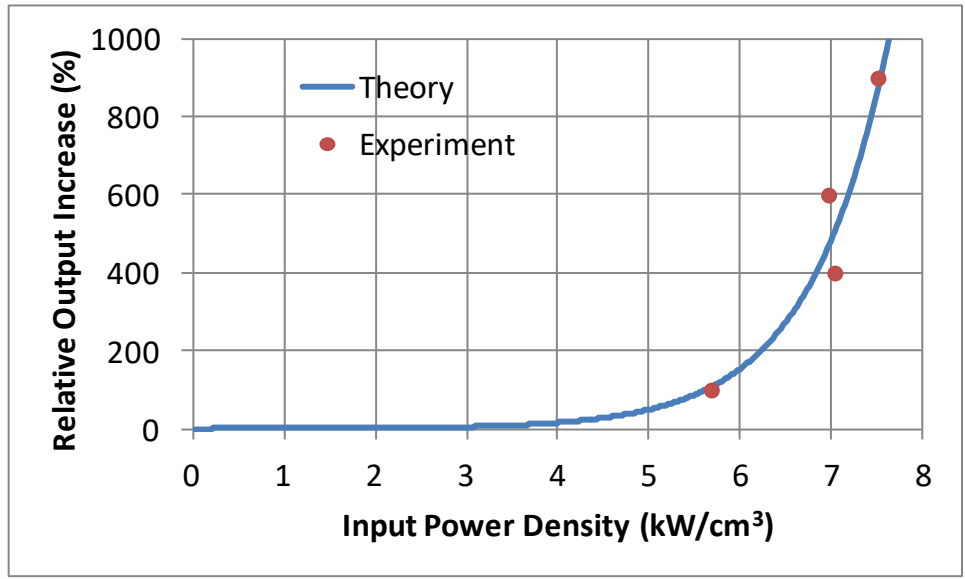

Figure 6: The output energy from p-$^{11}$B reactions exhibits threshold phenomena as a result of positive feedback in the fusion process and in good agreement with the observed percentage increase of energetic particles.

## Measurements of gain factor $G = P_{out} / P_{in}$

According to our concept, the power used to produce electrons and drive neutrals should be less than the power generated by fusion to prove the presence of fusion reactions.

Since our approach of lowering the fusion barrier results in a few fusion processes taking place simultaneously, an inclusive method is devised to operate the Alphas Ring system in a long-pulse mode (~20s) in which the increases in temperatures of all components were measured. Three regions of measurements, the central electrode, the outer electrode and the end plates have their own separate temperature sensors. The total output power is equal to the sum of all the heat generated in such a system. Because fusion events occur at the surface of the outer shroud, the largest contribution to the output heat comes from fusion reactions at this copper or tantalum shroud. The heat carried by water-cooling of the outer shroud is also included in the overall estimate of total heat generated.

The total input power is simply the electrical power supplied to the system from the wall plug. Presently the gain factor $G = P_{out}/P_{in}$ under several long-pulse conditions (t ~ 20 sec) averages to approximately 9. Results of four power extraction experiments are shown in Table 1a. The experimentally obtained power gain factor G as function of time in one long-pulse experiment is summarized in Table 1b.

Table 1a: Measured powers and calculated gain factors in four experiments using tantalum and copper shrouds. Output powers come mainly from the outer shroud where most fusion processes started.

| **Experiment #** | | | #1 | #2 | #3 | #4 |
|---|---|---|---|---|---|---|
| Outer shroud | Material | | Tantalum | Copper | Copper | Copper |
| | Specific heat | | 0.14 | 0.385 | 0.385 | 0.385 |
| | Mass (gm) | | 66 | 210 | 210 | 210 |
| | Max temperature | | 1600 | 92 | 106 | 324 |
| | Min temperature | | 27 | 12 | 16 | 80 |
| | Temp increase | | 1573 | 80 | 90 | 244 |
| | Time (sec) | | 1 | 0.5 | 0.5 | 1 |
| | Power out (kW) | | 14.5 | 12.9 | 14.5 | 19.7 |
| | | | | | | |
| Shroud cooling | Flow (cc/sec) | | | 29.6 | 25.5 | 0.33 |
| | Max temperature | | | 20 | 34 | 98.6 |
| | Min temperature | | | 12 | 16 | 28.25 |
| | Temp increase | | | 8 | 18 | 70.35 |
| | Power out (kW) | | | 0.9 | 1.9 | 0.1 |
| | | | | | | |
| Total output power | (kW) | | 14.5 | 13.8 | 16.4 | 19.8 |
| Total input power | (kW) | | 2 | 1.6 | 2.4 | 1.96 |

| Gain | (Output/Input) | 7.3 | 8.6 | 6.8 | 10 |

Table 1b: The gain factor is typically time-dependent because it takes time to heat up the electron emitters at a certain input power.

| Time (sec) | 1 | 2 | 16.5 | 17 |
|---|---|---|---|---|
| Output power from shroud cooling water (kW) | 0.002 | 0.002 | 0.096 | 0.096 |
| Output power from copper shroud surface (kW) | 0.686 | 1.37 | 21.3 | 19.7 |
| Output power from center electrode cooling (kW) | | 0.01 | 0.537 | 0.549 |
| Total output power (kW) | 0.688 | 1.38 | 21.8 | 20.25 |
| Total Input power (kW) | 0.72 | 2.56 | 2.52 | 1.96 |
| **Gain (output/input)** | **0.95** | **0.54** | **8.7** | **10** |

**Observation of self-sustaining reactions: evidence of Gain > 1**

With the gain factor G greater than unity we present a piece of evidence of self-sustaining reaction in an isolated $LaB_6$ emitter. As shown in Figure 7, an isolated piece of emitter, 400 microns in diameter, glows for 100 μsec and then splits into two equal pieces. This single piece observed with a telescopic lens was originally attached to the main piece of emitter. When it is broken off it becomes stationary in the stream of clockwise hydrogen neutrals and the counter-clockwise flow due to radial pressure gradient. As shown in Appendix F the number of fusion reactions can account for the energy required to split this piece of emitter into two equal halves.

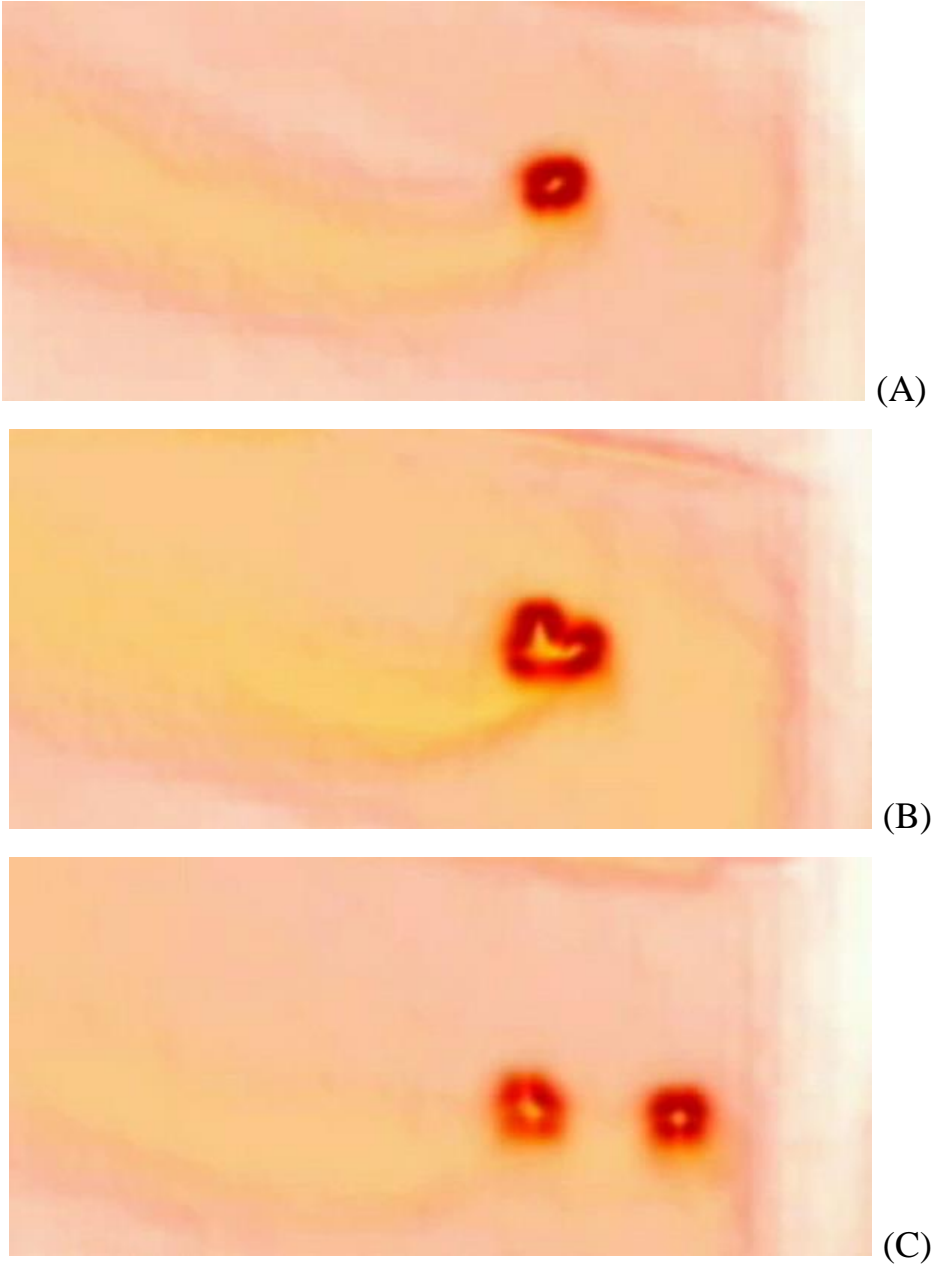

Figure 7: A single piece of LaB$_6$, 400μm across, splits up into two equal parts during interactions with hydrogen stream flowing from right to left. Pictures were taken with double He filters (587.5 nm). (A) A single piece of LaB$_6$ glows with a brighter core due to surface emission ($t = 0$); (B) The piece grows to twice the size ($t = 33$ μsec); (C) The single piece is split into two nearly equal pieces ($t = 100$ μsec).

**VII. General application of this approach**

This approach of utilizing high-density electrons and neutrals with positive feedback are not restricted to a rotating system. In fact, it is equally informative to perform experiments in nano and pico regimes where nuclear fusion reactions can be precisely monitored in the presence of collective electron dynamics. We are presently following this path of designing nano-systems [25-26] where fusion products can be diagnosed by surrounding alpha detectors. Both laser experiments and computer modeling are employed towards this goal.

The scaling of this concept to large systems has been examined. Since rotation does not depend on a mechanical rotor, an arbitrary large system as befits a power station can be developed. We have demonstrated the beneficial effects of electron screening and are now developing an important acceleration scheme such that a large number of neutrals at keV energy can be rotated. As a result, the fusion cross section will be much larger and the requirements of high-density electrons can be relaxed.

**Conclusion**

A distinguishing consequence of our approach is the attainment of more than break-even fusion. The large gain ratio of energy released compared to the incident energy is due to the fact that our approach relies on the proper positioning of electrons and neutrals and does not requires their energies to be high compared with the Coulomb barrier. The evidence of fusion has been confirmed by the appearance of MeV particles, energy balance, optical and mass spectra of helium and carbon atoms.

This approach should affect all types of fusion reactions. Collective electron dynamics play an important role in reducing the Coulomb barrier. The large concentration of neutrals has led to the positive feedback in fusion reactions, enabling aneutronic processes to dominate over neutronic processes.

The three features of our approach have been demonstrated experimentally and theoretically in a rotation system characterized by its large centrifugal acceleration. Other systems can also be used to achieve fusion at low energies and high densities when the collective electron dynamics is utilized. To our best knowledge, this is the first time fusion with positive gains is achieved in a table-top system under both pulse and steady-state conditions. Our approach can be scaled to larger or smaller systems for a wide range of applications.

**Appendix A: Coulomb potential energy between two hydrogen atoms**

To find out the screening energy of two interacting hydrogen atoms, we need to calculate the overall Coulomb energy between these two atoms as they approach each other. In the first order approximation, it can be assumed that the bound electron distributions are not affected by the approaching nuclei. The "atom-atom" potential energy between atoms 1 and 2 (with coordinates shown in Figure A1), can thus be calculated from the following integral,

$$U_{aa}(r) = e^2 \iint \frac{[\delta(\vec{r}_1)-n(\vec{r}_1)][\delta(\vec{r}_2)-n(\vec{r}_2)]}{|\vec{r}_1-\vec{r}_2|} (r_1^2 \sin\theta_1 dr_1 d\theta_1 d\varphi_1)(r_2^2 \sin\theta_2 dr_2 d\theta_2 d\varphi_2) \quad \text{(A1)}$$

$$|\vec{r}_1 - \vec{r}_2|^2 = r_1^2 + r_2^2 + r^2 + 2rr_1\cos\theta_1 - 2rr_2\cos\theta_2 - 2r_1r_2[\cos\theta_1\cos\theta_2 - \sin\theta_1\sin\theta_2\cos(\varphi_1 - \varphi_2)]$$

where $\delta(r)$ and $n(r)$ represent the point-charge nucleus and orbital electron density distribution separately. The ground-state electron density distribution in the hydrogen atom is given by 1s-state wave function ($a_0$ is the Bohr's radius)

$$n(r) = |\psi_{1s}(r)|^2 \quad ; \quad \psi_{1s}(r) = \frac{\exp(-r/a_0)}{\sqrt{\pi}a_0^{3/2}} \quad ; \quad a_0 = \frac{\hbar^2}{me^2} \quad \text{(A2)}$$

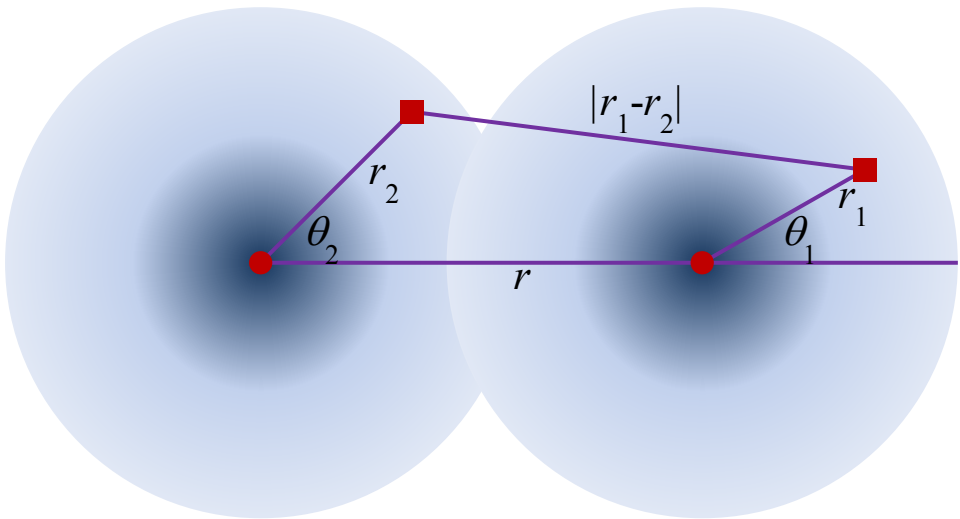

Figure A1: Coordinates for calculating potential energy between two hydrogen atoms.

In the first order approximation, the "atom-atom" potential energy can be seen as the combination of "atom-proton" and "atom-electron" potential energies.

$$U_{aa}(r) = U_{ap}(r) + U_{ae}(r) \tag{A3}$$

Instead of carrying out the tedious integral (A1) to obtain the "atom-proton" potential energy, $U_{ap}(r)$ can be solved readily from the Poisson's equation

$$\nabla^2 U_{ap}(r) = 4\pi e^2 [n(r) - \delta(r)] \quad ; \quad n(r) = \frac{\exp(-2r/a_0)}{\pi a_0^3} \tag{A4}$$

And the solution was found to be

$$U_{ap}(r) = e^2 \left(\frac{1}{r} + \frac{1}{a_0}\right) e^{-2r/a_0} \tag{A5}$$

which is exactly the "atom-proton" potential energy as shown in Eq. (13).

To obtain the "atom-electron" potential energy, we can take the "atom-proton" potential energy and integrate over the electron cloud density distribution (with negative charge).

$$U_{ae}(\vec{r}) = -\int U_{ap}(\vec{r} - \vec{r}_1) n(\vec{r}_1) d^3 \vec{r}_1$$

$$= -\frac{2e^2}{a_0^3} \int \left(\frac{1}{\rho} + \frac{1}{a_0}\right) e^{-\frac{2\rho}{a_0}} e^{-\frac{2r_1}{a_0}} r_1^2 \sin\theta_1 dr_1 d\theta_1 \tag{A6}$$

$$\rho^2 = r^2 + r_1^2 + 2rr_1 \cos\theta_1$$

Eq. (A6) can be integrated analytically using the transform, $\rho d\rho = rr_1 d\cos\theta_1$, and the result is

$$U_{ae}(r) = -\frac{e^2}{a_0} \left(\frac{3}{8} + \frac{3r}{4a_0} + \frac{r^2}{6a_0^2}\right) e^{-\frac{2r}{a_0}} \tag{A7}$$

By adding up the results in (A5) and (A7), we get the "atom-atom" potential energy as

$$U_{aa}(r) = U_{ap}(r) + U_{ae}(r) = \frac{e^2}{r}\left(1 + \frac{5r}{8a_0} - \frac{3r^2}{4a_0^2} - \frac{r^3}{6a_0^3}\right) e^{-\frac{2r}{a_0}} \tag{A8}$$

This is exactly the $r$-dependent potential energy shown in (16). For comparison, $U_{ap}$ and $U_{aa}$ are plotted in Figure A2 as function of nucleus separation. It is noted that "atom-proton" force is always repulsive, since the slope of the potential energy is always negative. However, the "atom-atom" potential energy has a minimum at $1.875a_0$. The force between 2 atoms is attractive if the separation is larger than this value and repulsive if smaller. It behaves similarly to the well-known Van der Waals force between two neutral molecules and is the result of electron cloud distributed extensively around the nucleus.

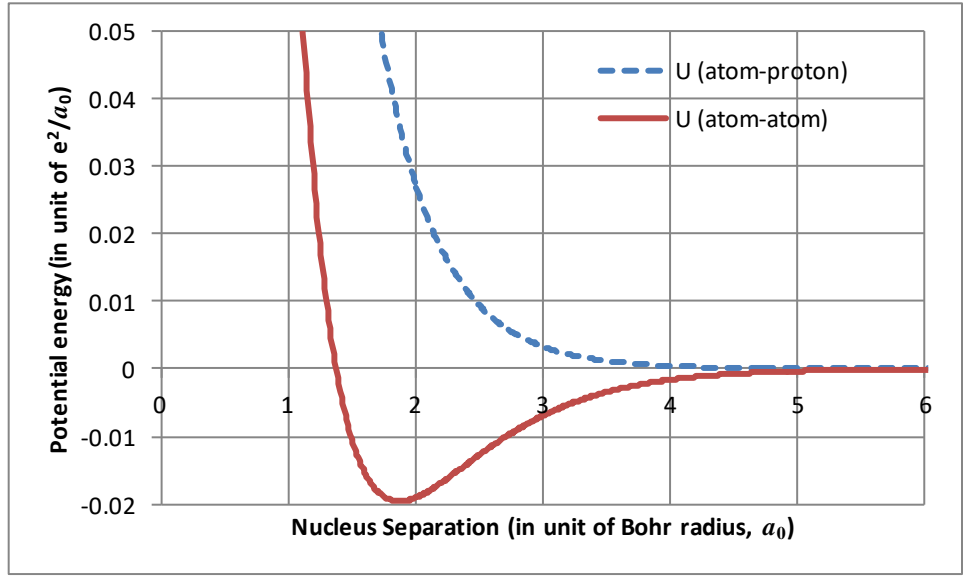

Figure A2: Coulomb potential energies between a neutral atom and a proton (blue dash) or between two atoms (red solid). It is noted there is a minimum in the atom-atom potential energy at $1.875a_0$. This is the result of electron cloud-cloud interaction and the force between two atoms turns attractive when they are separated farther than this distance.

**Appendix B: Threshold and saturation of the fusion process**

It has been shown in (23) that the screening energy (proportional to the plasma fluctuation amplitude) grows exponentially with time at the onset of fusion process,

$$E_s = E_{s0}[1 + q(e^{bt} - 1)] \tag{B1}$$

where $E_s = \dfrac{e^2 \Lambda^2}{\pi} n_1$ ; $b = \dfrac{3\beta\gamma e^2 \Lambda^2 v n_H n_B \sigma'(E_{s0})}{\pi}$ ; $q = \dfrac{\sigma(E_{s0})}{\sigma'(E_{s0}) E_{s0}}$

Note that $b$ is the exponential growth rate of $E_s$ and $q$ can be treated as a growth factor, which is a dimensionless parameter depending only on the intrinsic property of the particular fusion reaction and is a slowly varying function of $E_s$. Eq. (B1) describes only the start up phenomenon with exponential growth, but does not provide other interesting information, such as threshold conditions and steady-state operation.

*Threshold conditions*

There might be other processes with decaying power to overcome the exponential growth. In such cases, the fusion will grow very slowly or not at all. One of possible mechanisms is the decay of plasma wave oscillations. Including this effect, the equation for $E_s(t)$ becomes

$$\frac{dE_s}{dt} = b\left(E_s - E_{s0} + \frac{\sigma}{\sigma'}\right) - \alpha(E_s - E_{s0}) = (b-\alpha)(E_s - E_{s0}) + b\frac{\sigma}{\sigma'} \tag{B2}$$

where $\alpha$ is the damping rate of plasma waves. The solution of (B2) was found to be

$$E_s(t) = E_{s0}[1 + q\, f(x)bt] \text{ where } f(x) = e^x \frac{\sin x}{x} \text{ and } x = \frac{(b-\alpha)t}{2} \tag{B3}$$

Therefore, the threshold condition for starting the exponential growth is that the fusion growth rate has to be larger than the damping rate of plasma waves, i.e.

$$b > \alpha \quad \text{or} \quad 3\beta\gamma e^2 \Lambda^2 v n_H n_B \sigma'(E_{s0}) > \pi\alpha \tag{B4}$$

The function $f(x)bt$ in (B3) for the three regions ($b > \alpha$, $b = \alpha$, $b < \alpha$) are shown in Figure B1. It is interesting to note that, when $b = \alpha$, the screening energy (proportional to $n_1$) grows linearly in time, and as a result, the output power is expected to remain constant. If $b < \alpha$, the electron density fluctuation $n_1$ dissipates more quickly by plasma wave decay than grows by fusion reactions. As a result, the output power (proportional to the slope of the curve) decays with time.

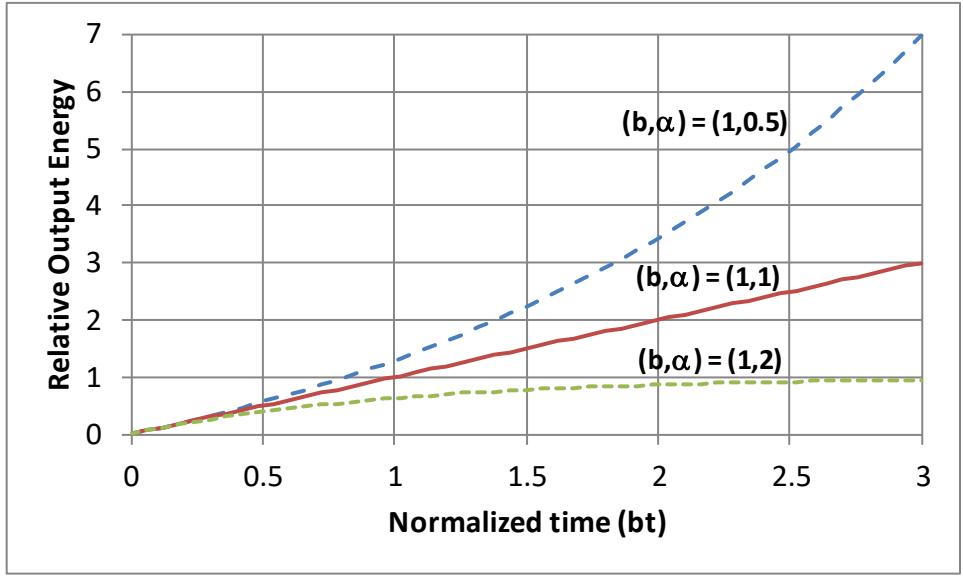

Figure B1: Comparison of Growths of screening energy (or, equivalently, plasma fluctuations) under three conditions: growth > dissipation (blue long dash); growth = dissipation (red solid); and growth < dissipation (green short dash).

*Saturation of fusion process*

On the other hand, to reach a steady state, the growth rate should decrease as $n_1$ grows. This saturation process could occur if the fractional plasma fluctuation decreases with increasing electron density. This is reasonable because, with limited supply of heat, the plasma wave fluctuations are not expected to grow without limit. We can use the following formula for the fractional fluctuation, $\gamma$, to model this saturation phenomenon

$$\gamma = \frac{n_s}{n_1+n_s}\gamma_0 \tag{B5}$$

where $\gamma_0$ is the small-$n_1$ fluctuation amplitude and $n_s$ is the saturation density. It can be seen $\gamma$ remains a constant when $n_1$ is small and gradually drops in value when $n_1$ becomes larger. By introducing Eq. (B5) in the model and neglecting the plasma wave decay, we can re-write the equation of $E_s$ as

$$\frac{dE_s}{dt} = \frac{b(E_s - E_{s0} + qE_{s0})}{1+\frac{E_s}{E_{ss}}} \quad \text{where} \quad \frac{E_s}{E_{ss}} = \frac{n_1}{n_s} \tag{B6}$$

and $E_{ss}$ is the saturated screening energy (equivalent to $n_s$ for density $n_1$). Eq. (B6) becomes

$$\left[1 + \frac{E_{ss} - (q-1)E_{s0}}{E_s + (q-1)E_{s0}}\right] \frac{dE_s}{E_{ss} dt} = b \tag{B7}$$

and can be integrated readily. The results are

$$\frac{E_s - E_{s0}}{E_{ss}} + \left[1 - \frac{(q-1)E_{s0}}{E_{ss}}\right] \ln\left(\frac{E_s + (q-1)E_{s0}}{qE_{s0}}\right) = bt \tag{B8}$$

If $E_{ss}$ is large, Eq. (B8) is reduced to the solution shown in (B1). If $(E_s - E_{s0})$ is large compared to $E_{ss}$, the first term on the left-hand side of (B8) dominates and we have

$$E_s \approx E_{s0} + E_{ss} bt \tag{B9}$$

which represents a linear increase in the output energy (i.e. constant output power). Eq. (B7) and (B8) can also be re-written in terms of the electron fluctuation amplitude $n_1$,

$$\left[1 + \frac{n_s - (q-1)n_{10}}{n_1 + (q-1)n_{10}}\right] \frac{dn_1}{dt} = bn_s \quad \text{and} \tag{B10}$$

$$\frac{n_1 - n_{10}}{n_s} + \left[1 - \frac{(q-1)n_{10}}{n_s}\right] \ln\left(\frac{n_1 - n_{10}}{qn_{10}} + 1\right) = bt \tag{B11}$$

The output power can then be derived from

$$P_{out} = VQ\kappa = \frac{VQ}{3\beta\gamma} \frac{dn_1}{dt} = \frac{VQp[n_1 + (q-1)n_{10}]}{3\beta\gamma_0} \approx \frac{VQpn_1}{3\beta\gamma_0} \tag{B12}$$

Assuming $V = 10^{-7}$ m³, $Q = 8.7$ MeV, $p = 10^2$ s⁻¹, $n_1 = 10^{25}$ m⁻³, $\beta = 10^5$, $\gamma_0 = 5 \times 10^{-3}$, we have

$$P_{out} = \frac{0.1 \times 8.7 \times 1.6 \times 10^{-13} \times 10^2 \times 10^{19}}{3 \times 10^5 \times 10^{-2}} = 23.2 \text{ kW} \tag{B13}$$

which is not far from what was observed for the steady-state output power in the lab. The gain factor can then be calculated as the ratio of output power to input power

$$G = \frac{P_{out}}{P_{in}} = \frac{P_{out}}{IV} \approx \frac{23.2 \text{ kW}}{2 \text{ kW}} \approx 11.6 \tag{B14}$$

where $I$ is the current flowing from the inner wall to the outer wall and $V$ is the voltage between these two walls. For the long pulse experiments, the input electric power is close to 2 kW. This calculated gain factor 11.6 is close to the measured gain values 10 as shown in Tables 1a & 1b.

To show the saturation process in a plot, we neglect the terms with $qn_{10}$ in (B11) and assume $n_s$ is much larger than $n_{10}$. Under these conditions, Eq. (B11) can be normalized as

$$\frac{n_1 - n_{10}}{qn_{10}} \to n_1 \quad ; \quad \frac{n_s}{qn_{10}} \to n_s \quad ; \quad bt \to t$$

$$\frac{n_1}{n_s} + \ln(n_1 + 1) = t \tag{B15}$$

The two terms on the left-hand side of (B14) describes two different time dependences. When the process is unsaturated, the second term dominates and represents an exponential growth in time. When the process is saturated, the first term takes over and results in a linear growth in time. The general behavior of the output power, which is proportional to the slope of $n_1(t)$, is shown in Figure B2 where the normalized saturation density, $n_s/n_{10}$, is set at 20. It can be seen that $n_1'$ (proportional to output power) increases exponentially at the beginning and is saturated eventually at a level close to $n_s$.

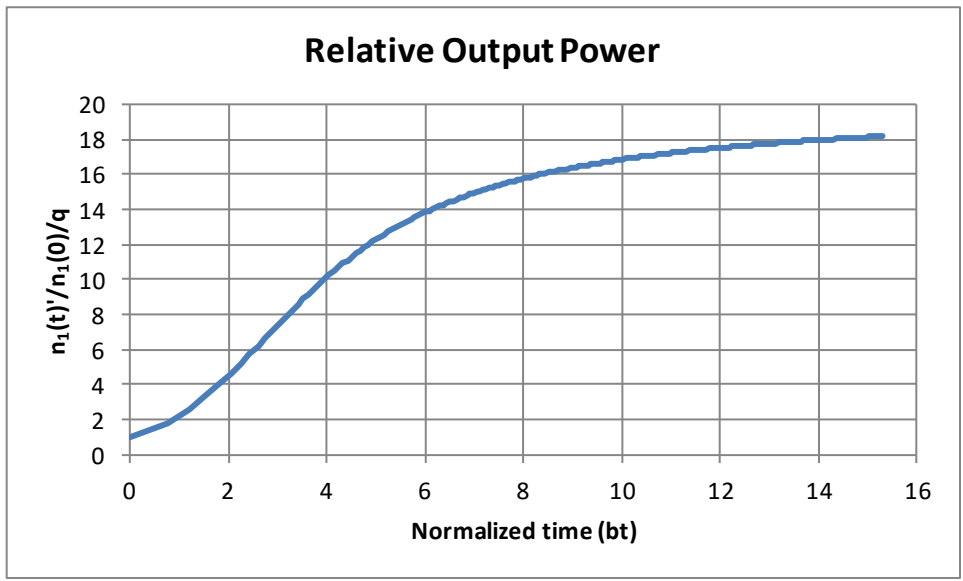

Figure B2: Growth and saturation of output power as function of time.

## Appendix C: Ion-neutral coupling

The following equations describe how neutrals through their frequent collisions with ions can be made to follow the external electromagnetic excitations.

$$Mn_i \frac{\partial v_i}{\partial t} = qn_i(\underline{E} + \underline{v}_i \times \underline{B}) - \nabla p_i + P_{ie} \tag{C1}$$

$$mn_e \frac{\partial v_e}{\partial t} = -qn_e(\underline{E} + \underline{v}_e \times \underline{B}) - \nabla p_e + P_{ei} \tag{C2}$$

where *M* and *m* are the mass of ion and electron, $n_{i,e}$ are the densities of ions and electrons, $v_{i,e}$ are the velocities of ions and electrons, $p_{i,e}$ are the momentums of ions and electrons, $\underline{E}$ is the electric field, $\underline{B}$ is the magnetic field, and the coupling term $P_{ie}$ is the momentum change due to ion-electron interaction (note that $P_{ie} = -P_{ei}$).

Neglecting the viscosity tensor and $(\underline{v} \cdot \nabla)\underline{v}$ terms and adding (C1) and (C2), the interaction terms are cancelled, and we have

$$\frac{\partial}{\partial t}(Mn_i \underline{v}_i + mn_e \underline{v}_e) = q[(n_i \underline{v}_i - n_e \underline{v}_e) \times \underline{B}] - \nabla(p_i + p_e) \tag{C3}$$

Equation (C3) can be written as

$$\rho \frac{\partial \underline{v}}{\partial t} = \underline{j} \times \underline{B} - \nabla p \tag{C4}$$

where $\rho = n_i M + n_e m = n(M + m)$ is the combined ion-electron density, and the average velocity and effective electric current are (assuming $n \sim n_i \sim n_e$).

$$\underline{v} = \frac{1}{\rho}(n_i M \underline{v}_i + n_e m \underline{v}_e) \approx \frac{M\underline{v}_i + m\underline{v}_e}{M+m} \tag{C5}$$

$$\underline{j} = q(n_i \underline{v}_i - n_e \underline{v}_e) \approx qn(\underline{v}_i - \underline{v}_e) \tag{C6}$$

Now, consider the interactions between the charge current and the neutrals. Their equations can be written as

Neutrals $\quad Mn_0 \frac{\partial v_0}{\partial t} = -\nabla p_0 + \underline{P}_{ni} \tag{C7}$

Charged $\quad \rho \frac{\partial \underline{v}}{\partial t} = \underline{j} \times \underline{B} - \nabla p_{ie} + \underline{P}_{in} \tag{C8}$

Adding (C7) and (C8), the interaction terms are cancelled and we obtain

$$Mn_0 \frac{\partial v_0}{\partial t} + \rho \frac{\partial \underline{v}}{\partial t} = \underline{j} \times \underline{B} - \nabla(p_0 + p_{ie}) \tag{C9}$$

Equation (C9) shows that the neutrals are influenced by $\underline{j} \times \underline{B}$ force as a result of collisions between charges and neutrals. Optical monitoring of Argon ions and neutrals in Figure C1 shows that they do rotate together and expand radially outward at the same rate. Another experiment on the linear acceleration of a combined mixture of neutrals and ions (Figure C2) also shows that they move together with an effective mass of N*m where N is greater than $10^6$.

Computer modeling and laboratory experiments with higher current drives and larger chamber diameters have shown that neutrals can be accelerated to keV range of energies. This implies much higher fusion cross sections than what can be achieved with our present rotating neutrals of energy in the range of 20 to 40 eV.

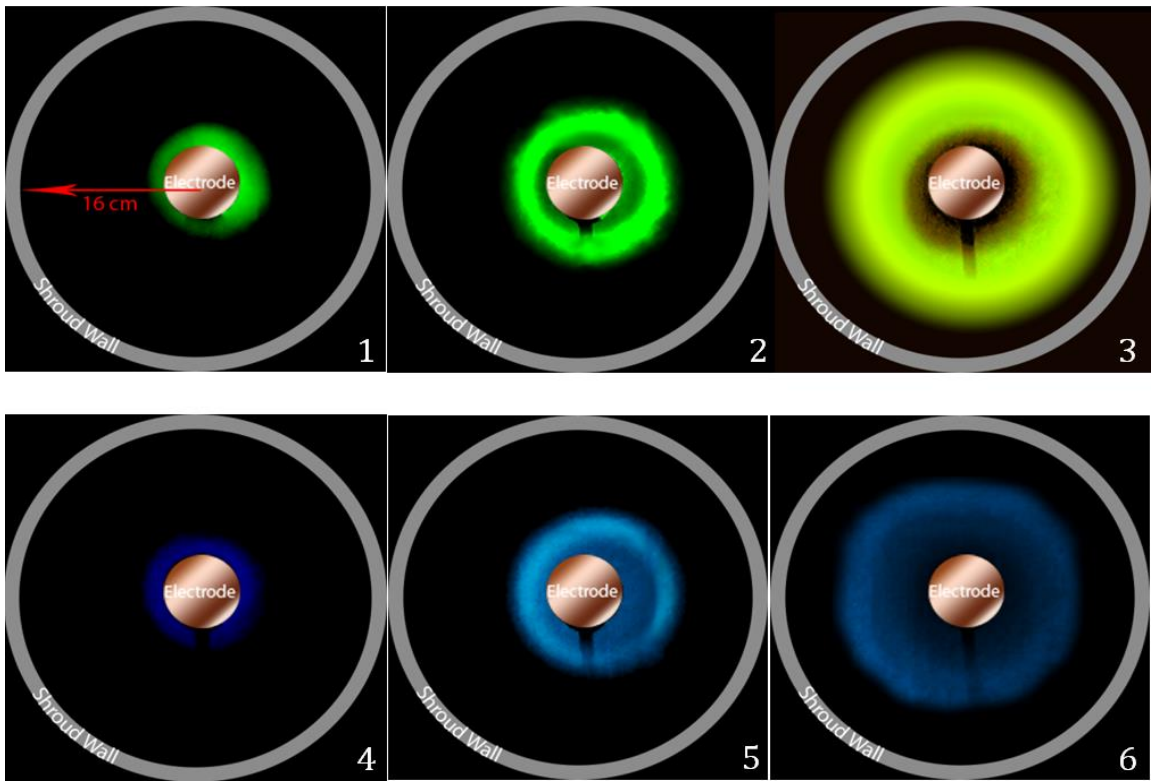

Figure C1: Spectroscopic measurements show that ions 442.6nm (blue) and neutrals 518.77nm (green) rotate and expand radially in the same manner.

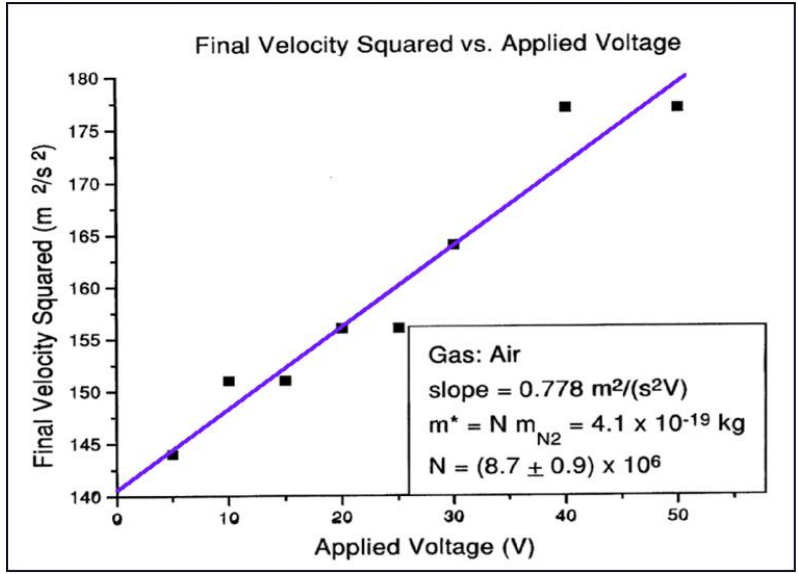

Figure C2: Experiments on linear acceleration of combined ions and neutrals show that as many as 8.7x10⁶ neutrals can be accelerated with each ion.

**Appendix D: Gas density enhancement factor**

In a rotating gas system, the centrifugal force on the neutrals is balanced by the gas radial pressure gradient when there is no gas flow in the radial direction,

$$F(r) = mn(r)r\omega^2 = \frac{dP}{dr} = kT\frac{dn(r)}{dr} \tag{D1}$$

Equation (D1) can be solved to get the $r$-dependent gas density

$$n(r) = n(0)\exp\left(\frac{\alpha r^2}{R^2}\right) \quad ; \quad \alpha = \frac{m\omega^2 R^2}{2kT} = \frac{E_R}{kT} \tag{D2}$$

where $R$ is the outer radius and $E_R$ is the kinetic energy at $R$. It is clear that the density has an exponential dependence on radius squared. Equation (D2) can be integrated over the space to obtain the average density $n_0$, which is equivalent to the initial uniform gas density before rotation. Using this value, the $r$-dependent density can be written as

$$n(r) = n_0 \alpha \exp\left(\frac{\alpha r^2}{R^2}\right)/(e^\alpha - 1) \approx n_0 \alpha \exp\left[\alpha\left(\frac{r^2}{R^2} - 1\right)\right] \tag{D3}$$

and $n(R) \approx n_0 \alpha$ if $\alpha \gg 1$. $\tag{D4}$

It has been demonstrated in the lab that the gas can be driven up to $3 \times 10^5$ RPS or higher without losing stability. Just for example, assume the gas is rotating at $10^5$ RPS and the radius is 10 cm. At this speed, the hydrogen molecule has a kinetic energy of 41.2eV. At room temperature ($kT = 0.026$eV), the value of $\alpha$ is about 1600. The gas density and the pressure near the outer wall are thus enhanced by this factor. For an initial gas pressure of 3 torr, the wall pressure at $10^5$ RPS will be about 6.3 atmospheres and the compressed gas density is $1.6 \times 10^{26}$ m$^{-3}$. This density is 6 orders of magnitude higher than typical hot ion plasma densities for the thermonuclear fusion.

## Appendix E: Locations of fusion echoes observed in Alpha Ring experiments

Echo process in general requires a distribution or collection of oscillators whose phases are dependent on certain spatial coordinates. In our fusion experiments we have a group of rotating neutrals whose angular rotation speeds depend on their radial locations. These neutrals pass through two electron emitters situated on the outer shroud with a separation angle of $\theta_0$.

For echoes to occur there must be oscillations of electric fields generated by dynamic temporal variations of electron densities at each source. Past experiments have shown that counter flows of ions and electrons are found to create oscillating double charge layers [19, 20]. As such oscillating electrons are carried in the azimuthal direction by the rotating neutrals, fusion processes occur when the oscillations of propagating electron densities from different sources arrive in phase at specific locations. Fusion process is identified by the emission of energetic ions whose orbits are influenced by the axial magnetic field. These "fusion echoes" demonstrate the evidence of oscillating electron densities and how they play important roles in the fusion process through creating higher densities and fields at certain locations and that there is a threshold for the reduction of the Coulomb barrier. The locations of echoes depend only on the distance between the two sources and the ratio of the oscillation frequencies of these two sources [21, 22].

Effect of perturbations by two locations 1 and 2 separated by an azimuthal distance of L = $r_0 \theta_0$. The summation of the oscillating fields contributed by the collection of rotating neutral streams with imbedded electrons of distribution $f(v)$ can be written as:

$$N_e(\theta, t) = \int f(v) e^{i\frac{\omega_1}{v} r_0 \theta - i\omega_1 t} e^{-i\frac{\omega_2}{v} r_0 (\theta - \theta_0) + i\omega_2 t} \, dv \tag{E1}$$

where $\omega_{1,2}$ are the plasma oscillation frequencies at locations 1 and 2. Echoes occur when the following phase factor becomes zero and all oscillations can contribute constructively,

$$\frac{\omega_1}{v} r_0 \theta - \frac{\omega_2}{v} r_0(\theta - \theta_0) = 0 \quad \text{or} \quad (\omega_2 - \omega_1)\theta = \omega_2 \theta_0 \tag{E2}$$

Echo angles $\theta$ can be calculated as follows:

$$\theta = \theta_0 \frac{\omega_2}{(\omega_2 - \omega_1)} \quad \text{or} \quad \theta = \theta_0 \frac{1}{\left(1 - \frac{\omega_1}{\omega_2}\right)} \tag{E3}$$

For example, if $\omega_2 = 2\omega_1$, we have $\theta = 2\theta_0$. If $\omega_2 = 3\omega_1$, we have $\theta = \frac{3}{2}\theta_0$. In general $\omega_1$ and $\omega_2$ can be harmonics of a fundamental frequency $\omega_0$ for high-order echoes. From (E3), we know it is necessary that $\omega_2 > \omega_1$ if the echoes are to occur in the forward direction and the ratio of two frequencies is a rational number, i.e. $\omega_2/\omega_1 = m/n$ and $m > n$ ($m$, $n$ are the harmonic numbers of $\omega_2$, $\omega_1$). In general, the higher-order echoes have a lower strength than the fundamental. All the observed echoing locations (the azimuthally separated bright spots in Figure 4) can be identified by choosing proper numbers of $m$ and $n$ or appropriate frequencies of $\omega_2$, $\omega_1$.

## Appendix F: Calculation of fusion reactions required to split the isolated piece of emitter

The following calculation shows that fusion reactions must have occurred to give the energy for the breakage of the floating piece of emitter.

Consider a piece of $LaB_6$ (400 microns in each dimension) floating in a gas stream but not physically connected to any electrode. It is at a temperature that it is emitting electrons. The energy W, required to split this piece into two equal parts within 100 usec, can be calculated from the flexural strength of $LaB_6$ which is equal to 200 MPa or $2 \times 10^8$ N/m². W is calculated to be 0.015 J by multiplying this strength by the volume of $LaB_6$. Assuming each fusion event yields 9 MeV and is 33% efficient or $5 \times 10^{-13}$ J towards contributing to the splitting energy, the number of required fusion reactions is $3 \times 10^{10}$.

We now show that the number of fusion reactions $N$ taking place in 100 μsec within the volume of the floating piece of $LaB_6$ is comparable to this required number. This number $N$ is calculated from the fusion rate $dN/dt$ times the time interval and volume:

$$N = dN/dt\, \Delta t V = N_p N_b\, \sigma v \Delta t V = 5\times 10^{25}/\text{s·m}^3\, 7\times 10^{-11} \times 10^{-4} = 3\times 10^{10} \qquad (F1)$$

where $\sigma = 10^{-35}$ m$^2$ has been used, which corresponds to $E_s \sim 28$ keV or $n_1 \sim 6\times 10^{25}$ m$^{-3}$. It is interesting to note that the fusion reactions in this small piece are more active than other places and reach the critical electron density fluctuation earlier to cause the breaking up.

**Acknowledgments**

This research was entirely supported by the Alpha Ring Inc. Research Funds. We wish to thank our scientific and technical colleagues, J. Chen, Dr. L.C. Lee, Dr. K.H. Lee, Z. Lucky, Dr. R. Li, Dr. J. T. Cremer, Dr. A. Barzilov, Dr. C. Brown, S. Adimadhyam and S. Basu. The encouragement and support from L. Ng, A. Kwok and Dr. M. Lee have been very essential to the progress of this work.